\documentstyle[12pt,epsf,epsfig,amsmath]{article}
\def\eqright #1\cr{\noalign{\hfill$\displaystyle{{}#1}$}}
\def\eqleft #1\cr{\noalign{\noindent$\displaystyle{{}#1}$\hfill}}

\def\oldreffmt#1{\rlap{[#1]} \hbox to 2\parindent{}}

\def\figfmt#1{\rlap{Figure {#1}} \hbox to 1in{}}

\def\sectioneq{\def\theequation{\thesection.\arabic{equation}}{\let
\holdsection=\section\def\section{\setcounter{equation}{0}\holdsection}}}%

\newcounter{holdequation}

\def\begineq #1\endeq{$$ \refstepcounter{equation}\eqalign{#1}\eqno
	(\theequation) $$}
\def\contlimit{\,{\hbox{$\longrightarrow$}\kern-1.8em\lower1ex
\hbox{${\scriptstyle (a\rightarrow0)}$}}\,}
\def\centeron#1#2{{\setbox0=\hbox{#1}\setbox1=\hbox{#2}\ifdim
\wd1>\wd0\kern.5\wd1\kern-.5\wd0\fi
\copy0\kern-.5\wd0\kern-.5\wd1\copy1\ifdim\wd0>\wd1
\kern.5\wd0\kern-.5\wd1\fi}}
\def\centerover#1#2{\centeron{#1}{\setbox0=\hbox{#1}\setbox
1=\hbox{#2}\raise\ht0\hbox{\raise\dp1\hbox{\copy1}}}}
\def\centerunder#1#2{\centeron{#1}{\setbox0=\hbox{#1}\setbox
1=\hbox{#2}\lower\dp0\hbox{\lower\ht1\hbox{\copy1}}}}
\def\lsim{\;\centeron{\raise.35ex\hbox{$<$}}{\lower.65ex\hbox
{$\sim$}}\;}
\def\gsim{\;\centeron{\raise.35ex\hbox{$>$}}{\lower.65ex\hbox
{$\sim$}}\;}

\def\super#1{\ifmmode \hbox{\textsuper{#1}}\else\textsuper{#1}\fi}
\def\textsuper#1{\newcount\holdspacefactor\holdspacefactor=\spacefactor
$^{#1}$\spacefactor=\holdspacefactor}

\def\getcite#1,{\advance\citenumber by1
\def\getcitearg{#1}\def\lastarg{@}
\ifnum\citenumber=1
\ref{#1}\let\next=\getcite\else\ifx\getcitearg\lastarg\let\next=\relax
\else ,\ref{#1}\let\next=\getcite\fi\fi\next}

\def\pom{{\rm P\kern -0.53em\llap I\,}}
\def\spom{{\rm P\kern -0.36em\llap \small I\,}}
\def\sspom{{\rm P\kern -0.33em\llap \footnotesize I\,}}

\relax
\def\contlimit{\,{\hbox{$\longrightarrow$}\kern-1.8em\lower1ex
\hbox{${\scriptstyle (a\rightarrow0)}$}}\,}
\def\upon #1/#2 {{\textstyle{#1\over #2}}}
\relax
\renewcommand{\thefootnote}{\fnsymbol{footnote}}

\sectioneq

\def\til#1{\centeron{\hbox{$#1$}}{\lower 2ex\hbox{$\char'176$}}}
\def\tild#1{\centeron{\hbox{$\,#1$}}{\lower 2.5ex\hbox{$\char'176$}}}
\def\sumtil{\centeron{\hbox{$\displaystyle\sum$}}{\lower
-1.5ex\hbox{$\widetilde{\phantom{xx}}$}}}

\begin{document} 

\begin{titlepage} 

$~$

\vspace{1in} 

\begin{center} 
  
{\large\bf Getting to Know BFKL}

\medskip

Alan. R. White\footnote{arw@hep.anl.gov }

\vskip 0.6cm

\centerline{Argonne National Laboratory}
\centerline{9700 South Cass, Il 60439, USA.}
\vspace{0.5cm}

\end{center}

\begin{abstract}

 As a 30th anniversary tribute, I discuss the present and 
possible future impact of the physics and the physicists of BFKL.

\end{abstract} 

\vspace{1.5in}

\centerline{Contribution  to the 
Proceedings of HSQCD 2005 (recently requested).}

\renewcommand{\thefootnote}{\arabic{footnote}} \end{titlepage}

\section{Introduction} 

The BFKL equation was first derived thirty years
ago and at this meeting we have a special session dedicated to the
anniversary\footnote{As is common, for most of
my discussion I will not distinguish between the original 
derivation by Fadin, Kuraev, and Lipatov in a spontaneously-broken gauge theory
and the direct application to QCD based on 
later work of Balitsky and Lipatov that added the B to FKL.}. In this talk,
I would like to record a personal tribute to the achievement and 
impact of the physics and the physicists involved. I will do so via my own 
version of ``getting to know BFKL'' (which includes well-known happenings).
I will also describe a possible long-term significance which is very different
from present-day applications of the BFKL equation. 

I have known and interacted
with BFKL almost since the birth of the equation. In particular,
Lev Lipatov has been both a major influence and catalyst for my own research 
and a good friend, over much of the time period. For many people Lipatov is the obvious 
descendant of the grand line of russian physicists that passes through Landau, Pomeranchuk
and Gribov and, indeed, my first interaction with 
Lev and his collaborators was during an extended visit with
Vlodya Gribov. To present my own perspective on the significance 
of BFKL, it will be helpful to first talk about my early research and also talk about
my interactions with Gribov.

\section{Before BFKL}

During my early years in physics, 
immersed in regge theory and S-Matrix theory in Cambridge, I acquired
a lasting admiration of russian physics. As graduate students, Peter Goddard and I 
worked together to master the depths of Toller's 
group-theoretic multi-regge formalism. As an off-shoot, we understood
how complex helicity should be handled and this led me,
as a post-doc, to the remarkable 
paper by Gribov, Pomeranchuk, and Ter-Martirosyan (GPT) deriving reggeon 
unitarity. Even though the treatment of complex helicity was a problem, 
as I had been told, I was astonished by 
the spectacular leap made from low order field theory calculations
to general discontinuity formulae.

At Fermilab, in the Reggeon Field Theory (RFT) group of 
John Bronzan, Bob Sugar, Henry Abarbanel and Jochen Bartels (a life-long
friend and true BFKL afficionado), admiration of russian physics was
a communal affair. I came away with the Critical Pomeron heavily imprinted on my psyche,
even though I argued against it so strongly at first that I missed the
opportunity to be on the historical paper.
I went on to Berkeley, where Henry Stapp and I developed
the general S-Matrix dispersion theory needed to 
properly derive the multiparticle complex angular momentum and helicity
theory underlying the GPT paper. It has since become apparent to me
that this paper was years (perhaps decades) before it's time\footnote{It is barely
mentioned in ``The Analytic S-Matrix'' (the Cambridge bible of the time)
with the reggeon unitarity formula referred to as heuristic.}. Combining 
the beautiful mesh of multi-regge behavior 
and  multiparticle analyticity properties (that Henry Stapp and I saw) with the GPT paper,
convinced me that the unitarity of an S-Matrix must surely involve regge pole behavior 
in a fundamental manner - with the unitary Critical Pomeron as an essential ingredient.

\section{Meeting Gribov}

I first met Gribov at the 1977 EPS meeting in Budapest. (His first
conference, with significant numbers of western physicists present, 
in over a decade.) For   
most of the world it was Gribov's talk on  ``Non-abelian Gauge Copies'' 
that made this a historic meeting. For me it was the giddy experience of
learning that Gribov endorsed the supercritical RFT
that I had developed. This was a 
major controversy and I was almost 
alone  (as usual) in a minority viewpoint - insisting on reggeon unitarity! 
It is hard to exagerate the boost that Gribov's
support gave me. After the meeting was over, I met with Gribov for discussion and
began planning a visit to Leningrad as soon as I returned to CERN.

\section{Leningrad in 1978 - meeting FKL}

I visited Leningrad for six weeks (with a week in Moscow)
during the winter of 1978. It was bitterly cold and I could
fill a book with stories of my experience of communist
russia (part of the time with my wife and young son). I met often with Gribov
to discuss supercritical RFT and I gave ``Leningrad talks'' on this subject and on 
multiparticle complex angular momentum and dispersion theory. In our discussions,
it was hard to divert Gribov
from talking (to me) about anomalies. I must have 
absorbed something because, 
subsequently, anomalies have become a dominant part of my own thinking.

A very memorable event occured in the middle of my stay. Lev Lipatov suggested
we meet for a discussion. Subsequently, three very burly and physically overwhelming 
russians arrived to meet me for lunch in my hotel restaurant.
Victor Fadin, Eduard Kuraev and Lev Lipatov were doing
me the honor of coming, in full force, to explain the BFKL equation to me.
The downing of glasses of vodka that preceded lunch was very unfamiliar to me, as was the
physics I heard afterwards. With my narrow upbringing in S-Matrix theory
(and vodka inexperience), I was 
ill-equipped to properly appreciate what I was being told. Only later would I understand 
the fundamental and deeply insightful use of unitarity and analyticity
that would, ultimately, draw me to BFKL.

\section{Beginning to Study BFKL}

Shortly after returning from Leningrad I realized that my supercritical RFT 
contained a vector reggeon (exchange-degenerate with the pomeron), suggesting it should be
related to a spontaneously-broken gauge theory. I wanted 
to learn about regge behavior in gauge theories - rapidly! From the outset,
I was not content with finite-order perturbative calculations but rather
wanted to see the impact of unitarity directly.
Most of all, of course, I wanted to locate the Critical Pomeron via my supercritical RFT.

I liked the Grisaru and Schnitzer argument that reggeization
is produced by a non-abelian symmetry group and t-channel elastic unitarity.
Also impressive, were the Bronzan and Sugar papers demonstrating that, in agreement with GPT,
the Cheng and Lo results could be written in terms of reggeon diagrams
(with the eighth and tenth order results, in effect, predicted by the
sixth-order results).
However, the Cheng and Wu technique of tracking large light-cone 
momenta through diagrams seemed complicated to utilise and lacking 
insight into the role of unitarity. Fortunately, Jochen Bartels had arrived at CERN
and was developing his own program, based entirely on unitarity. He  
was an invaluable consultant and guide to the, often less accessible,
russian literature.

My appreciation of the unitarity plus dispersion relation 
methods utilized by BFKL (explained to me, presumably, in Leningrad)
was immediate. The simplicity of the born level calculation of gluon-gluon scattering 
was stunning (here ``gluons'' have a mass
generated by the Higgs mechanism.)
Renormalizability (or, more indirectly, unitarity boundedness) 
implies that the leading high-energy term satisfies
an unsubtracted t-channel dispersion relation.
The only singularity in the $t$-channel comes from
the pole diagram. Consequently, 
the full result is obtained from a single diagram. 
The summation of many diagrams is completely by-passed by a simple
exploitation of unitarity and analyticity! 

The calculation of higher orders is equally simple.
Multi-regge tree amplitudes can also be calculated by
using unsubtracted dispersion relations in the 
t-variables and renormalizability bounds.
Using s-channel unitarity and another dispersion relation then
gives the leading log amplitude to all orders. Hundreds of feynman diagrams are 
effectively summed and yet the result, by now extremely
well known, is extraordinarily simple, i.e gluons (and quarks) become regge poles
with a calculated trajectory function\footnote{McCoy and Wu require several volumes of 
the Physical Review to arrive at the simple conclusion that, in QED,
the leading log result, up to twelfth-order, is that the electron reggeizes.}  
Continuing the argument, BFKL show that when exchanged gluons are replaced
by reggeized gluons the unitarity calculation again
gives back the reggeized gluon as the leading result. 
This remarkable ``bootstrap'' shows that, in leading logs, reggeized gluon
exchange satisfies full s-channel unitarity
and elastic t-channel unitarity\footnote{Bartels has also extended the 
bootstrap to various multiparticle amplitudes.}. This is, essentially,
the ``multiperipheral bootstrap'' that many people hoped the pomeron would satisfy. 

The non-leading amplitudes contain the exchange of two
reggeized gluons. At first sight, reggeon unitarity is satisfied with the (massive)
BFKL kernel as the two reggeon interaction and 
a beautiful solution of both s-channel unitarity and reggeon unitarity 
is emerging. Unfortunately, reggeon unitarity is surely 
spoiled by the vacuum channel appearance of the BFKL pomeron,
due to the large transverse momentum scaling of the kernel. 
Reggeon unitarity requires the fundamental j-plane singularities to be 
regge poles. If they are not, 
it is hard to see how multiparticle t-channel
unitarity can be satisfied. Although the large transverse momentum scaling 
is essential for the appearance of the BFKL pomeron in QCD, reggeon unitarity
would be saved if this scaling instead produced 
an infra-red divergence related to a supercritical condensate.
How this relates to QCD is described in my 
companion talk. It would be some time before I was able to explain to Lev Lipatov the
relationship between his pomeron and my condensate.

\section{The Fall of the Iron Curtain and HERA}

During most of the 80's BFKL, like most other 
russian physicists, remained locked in 
the Soviet Union. Bartels and I were amongst the few westrn physicists 
referencing BFKL and the authors were not directly working on the subject.
After Leningrad, I did not meet Lipatov again until the 1987 
Protvino meeting (which was vodka-free - thanks to Gorbachev). Shortly 
afterwards the collapse of the Soviet Union began and russian
physicists started to appear everywhere in the west. Lev Lipatov visited CERN in 1988 and
in 1989 Victor Fadin (and Misha Ryskin) came to the Blois 
conference in Chicago that I organized with Marty Block. Even before HERA began,
Jochen Bartels had the foresight to see how important BFKL would 
become. He began organizing workshops at which BFKL would appear (in various combinations)
and later, as HERA was well underway, extended BFKL visits to DESY would become the norm. 
(Lev Lipatov would be the first to tell me that H1 had a structure function 
looking very like my pomeron!)

\section{The 90's and NLO BFKL}

Following Jochen's success at HERA, I started a series of Argonne/Fermilab workshops, with
Mike Albrow, at which  
BFKL would also appear in various combinations. At one workshop we were supposed to have
B, F \& L but, due to an airport mix-up in Moscow, Fadin 
did not arrive and Lipatov was forced to give five lectures. Lev also made 
several extended visits to Argonne, much to my delight. BFKL 
became an established part of extended perturbative QCD. 
Although there were qualifications, it became generally accepted that the BFKL pomeron
should appear in a variety of small-x processes and it was a constant 
source of experimental comparisons, even if a distinctive, clear, sighting proved elusive.
Many theorists and phenomenologists worked on the subject and citations accumulated
rapidly, as they continue to do.

For Victor Fadin and Lev Lipatov the 90's were a highly productive period. They were often
able to work together in institutions where they were well supported
and had the freedom to work intensely. The results were an achievement of major proportions.
They were able to complete the calculation of the NLO BFKL kernel and demonstrate that
it retained all the attractive properties of the leading-order kernel. In addition it was
shown that reggeization persists and (more recently and also, perhaps, even more remarkably)
that the bootstrap condition is still satisfied. Since unitarity, in all possible channels,
and sophisticated dispersion relation techniques were used, it would be very difficult
to determine just how large the enormous number of feynman diagrams involved 
actually is. In fact for some field theorists the techniques of Fadin and Lipatov
are sufficiently bewildering that they find the result difficult to accept. With my background
in unitarity and analyticity, the results are overwhelmingly impressive.
   
In the midst of all the BFKL excitement I (temporarily) abandoned my 
supercritical pomeron focus to see how much of BFKL I could derive from 
multiparticle j-plane unitarity in the t-channel. I succeeded in deriving the 
leading-order BFKL kernel, the (so-called) triple pomeron vertex, and also
a NLO kernel. Working with Claudio Coriano, I was able to prove that this NLO kernel
was conformally invariant and to derive the eigenvalue spectrum. This kernel
has subsequently been identified by Kirschner as part of the full BFKL kernel
and it has also been shown to fit experiment particularly well. Nevertheless, I have not
tried to determine explicitly how my derivation relates to the Fadin and Lipatov 
derivation. With Mark Wusthoff and Claudio, we were able to express the conformal
invariance of my kernel in a compact logarithmic form. Mark also found analagous, 
rather beautiful, much higher-order kernels, that he never published.

\section{The future }

Going signicantly beyond the NLO BFKL equation, 
with the aim of obtaining a fully unitary theory,
is an extremely difficult challenge. Lev Lipatov and Ian Balitsky have separately developed
very different effective action approaches. 
Unfortunately, neither approach 
seems to offer any hope for finding critical behavior involving a finite number of interactions 
or degrees of freedom. Without a transition to pomeron regge pole degrees of freedom, via
some form of confinement,
gluon infra-red behavior inevitably makes arbitrarily high-order interactions
equally important.

In my companion talk, I have described how, in a special version of QCD
(and a very special SU(5) theory), massless fermion 
chiral anomalies within the bound-state S-Matrix create an additional divergence which
produces confinement and a regge pole pomeron .
I use supercritical RFT to argue that there is 
a critical phenomenon involving dynamical infra-red chirality transitions
and infra-red scaling gluon reggeon kernels that produces
the unitary Critical Pomeron.
The bound-state S-Matrix is obtained from
multi-regge amplitudes that require the full armory of BFKL reggeon
diagrams and reggeon unitarity for their construction.
 
If LHC results were to steer physics in my direction, the calculation of 
bound-state amplitudes via reggeon diagrams would, very likely, become the focus 
of much of particle physics. In this case, 
the fundamental establishment of the reggeon diagram formalism, by BFKL, 
would have a long term significance in a manner, and in applications, very different 
from it's current success.

\end{document}